\documentclass[a4paper,fleqn,usenatbib]{mnras}
\usepackage{newtxtext,newtxmath}

\usepackage[T1]{fontenc}
\usepackage{ae,aecompl}

\usepackage{ifpdf}

\usepackage{graphicx}	
\usepackage{amsmath}	
\usepackage{amssymb}	
\usepackage{scalefnt}

\def\msun{{\rm M}_\odot}
\def\lsim{\mathrel{\rlap{\lower 3pt \hbox{$\sim$}} \raise 2.0pt \hbox{$<$}}}
\def\gsim{\mathrel{\rlap{\lower 3pt \hbox{$\sim$}} \raise 2.0pt \hbox{$>$}}}

\newcommand{\comments}[1]{} 

\title[Bar resilience to flybys]{Bar resilience to flybys in a cosmological framework}

\author[T. Zana et al.]{Tommaso Zana,$^{1}$\thanks{E-mail: tzana@studenti.uninsubria.it} Massimo Dotti,$^{2,3}$ Pedro~R. Capelo,$^{4}$ Lucio Mayer,$^{4}$ \newauthor Francesco Haardt,$^{1,3}$ Sijing Shen$^{5}$ and Silvia Bonoli$^{6}$\\
$^{1}$DiSAT, Universit\`a degli Studi dell'Insubria, Via Valleggio 11, IT-22100 Como, Italy\\
$^{2}$Dipartimento di Fisica G. Occhialini, Universit\`a di Milano-Bicocca,
Piazza della Scienza 3, IT-20126 Milano, Italy\\
$^{3}$INFN, Sezione di Milano-Bicocca, Piazza della Scienza 3, IT-20126 Milano,
Italy\\
$^{4}$Center for Theoretical Astrophysics and Cosmology, Institute for 
Computational Science, University of Zurich,\\
Winterthurerstrasse 190, CH-8057 Z$\ddot{u}$rich, Switzerland\\
$^{5}$Institute of Theoretical Astrophysics, University of Oslo, Postboks 1029, 0315 Oslo, Norway\\
$^{6}$Centro de Estudios de F{\'i}sica del Cosmos de Arag{\'o}n, plaza San Juan, 1 planta-2 ES-44001 Teruel, Spain
}

\date{Accepted: 2018 July 06; Revised: 2018 May 04; Received: 2018 July 03}

\pubyear{2018}

\begin{document}

\label{firstpage}

\pagerange{\pageref{firstpage}--\pageref{lastpage}}

\maketitle


\begin{abstract}
It has been proposed that close interactions with satellite galaxies can significantly perturb the morphology of the main galaxy. However, the dynamics of an already formed bar following the interaction with the external environment has not been studied in detail in a fully cosmological context. In this work, analysing the cosmological zoom-in simulation Eris2k, we study the effects that a very unequal-mass flyby crossing the stellar disc has on the stability of the pre-existing bar. We characterize the evolution of the bar strength and length showing that the perturbation exerted by the flyby shuffles the orbits of stars for less than one Gyr. After this time, the bar shows a remarkable resilience, reforming with properties comparable to those it had before the interaction. Our work shows that close unequal-mass encounters, the most frequent interactions occurring during the evolution of cosmic structures, have (i) an overall minor impact on the global evolution of the bar in the long term, still (ii) the effect is destructive and (iii) a very weak interaction is sufficient to dismantle a strong bar leading to its ``apparent death''. As a consequence, due to the non-negligible duration of the bar-less period, a fraction of observed spiral galaxies classified as non-barred could be prone to bar formation.
\end{abstract}



\section{Introduction}\label{sec:introduction}

Bars are common structures, being observed in more than one third of late-type galaxies in the local Universe, with the local bar fraction increasing with the galactic stellar mass \citep[see, e.g.][and references therein]{Consolandi_2016}. Their formation process is, notwithstanding, still debated: bars may form in isolation from the growth of small internal non-axisymmetries \citep[e.g.][]{Hohl_1971, Ostriker_Peebles_1973, Sellwood_2014}, or they may be triggered by tidal interactions with external structures, such as flybys and mergers, expected to occur in large numbers in the currently accepted hierarchical model \citep[e.g.][]{Byrd_et_al_1986, Mayer_Wadsley_2004, Curir_et_al_2006, Gauthier_et_al_2006, Romano-Diaz_et_al_2008, Martinez-Valpuesta_et_al_2016, Peschken_Lokas_2018}, or whole clusters \citep[e.g.][]{Byrd_Valtonen_1990, Lokas_et_al_2016}. The relative importance of these two formation channels is largely unconstrained due to the enormous computational burden that, to date, has limited the number of cosmological simulations with a resolution high enough to follow the growing instability \citep{Romano-Diaz_et_al_2008, Kraljic_et_al_2012, Scannapieco_Athanassoula_2012, Goz_et_al_2015, Okamoto_et_al_2015, Algorry_et_al_2017, Sokolowska_et_al_2017, Spinoso_et_al_2017}. Further uncertainty is due to the fact that interactions can either promote or delay the bar formation process, depending on the internal and orbital details of the encounter \citep[e.g.][]{Moetazedian_et_al_2017, Pettitt_Wadsley_2018, Zana_et_al_2018}.

Here we contribute to the study of the effect of external interactions on to bars, by focussing on the impact that cosmologically motivated flybys have on already developed stellar bars. To do so, we analyse the cosmological zoom-in simulation Eris2k \citep[][]{Sokolowska_et_al_2016,Sokolowska_et_al_2017} featuring state-of-the-art prescriptions for baryonic physics (see Section~\ref{sec:simulation}). A disc galaxy forms in the centre of the high-resolution region, showing a strong stellar bar extending over more than 5~kpc at $z < 1$. Within the Eris suite (e.g. Eris, \citealt{Guedes_et_al_2011}; ErisLE, \citealt{Bird_et_al_2013}; ErisBH, \citealt{Bonoli_et_al_2016}; etc.), the Eris2k run has the most developed bar both in strength and in length. A deeper comparison between Eris2k and ErisBH (the only other run with a discernible stellar bar; \citealt{Spinoso_et_al_2017}) is the topic of Zana et al., in preparation. The exquisite mass resolution of Eris2k is of paramount importance for this study, since it allows to analyse the effect of a very unequal-mass flyby passing well within the main stellar disc. Such close interactions have been proposed to be the strongest perturbations to the evolution of bars in isolation \citep{Moetazedian_et_al_2017}. A coarser resolution would indeed strongly underestimate the number density of small structures which, being more common, have a higher (non-negligible) chance to experience close pericentric passages at low redshift, after the initial phase of bar growth.

Some works have investigated in detail the effect of tidal perturbations on already-barred galaxies in isolated simulations and more simplified contexts \citep{Gerin_et_al_1990, Sundin_Sundelius_1991, Sundin_et_al_1993}. We try here to generalize the results in a more realistic environment (i.e. in a cosmological background), also taking advantage of the optimal spatial and time resolution provided by the Eris2k run. We also note that the response of barred systems to gravitational perturbations has been recently analysed \citep{Peschken_Lokas_2018} in a large sample of simulated galaxies using the Illustris-1 cosmological simulation \citep{Vogelsberger_et_al_2014}. However, we wish to reiterate that works based on cosmological volumes are very valuable to study the demographics and general properties of barred galaxies, but the much higher resolution (of order 0.1~kpc) possible with cosmological zoom-in simulations of individual galaxies is necessary to analyse in detail the internal dynamics of bars and their resulting evolution.

The analysis of the effect of the interaction and its comparison with previous works is presented in Section~\ref{sec:results}, whereas our conclusions are discussed in Section~\ref{sec:conclusions}.


\section{The simulation}\label{sec:simulation}

Eris2k is a cosmological zoom-in simulation of a Milky Way-sized galaxy, performed in a Wilkinson Microwave Anisotropy Probe three-yr cosmology ($\Omega_{\rm M} = 0.24$, $\Omega_{\Lambda} = 1 - \Omega_{\rm M}$, $\Omega_{\rm b} = 0.042$, $h = 0.73$, $n = 0.96$, and $\sigma_8 = 0.76$; \citealt{Spergel_et_al_2007}) and run with the $N$-body, smoothed particle hydrodynamics (SPH) code {\textsc{gasoline}} \citep[][]{Stadel_2001, Wadsley_et_al_2004} down to $z = 0.3$. The galaxy was chosen from a low-resolution, dark-matter (DM)-only run in a cosmological box of size (90~cMpc)$^3$ and re-run within a Lagrangian volume of (1~cMpc)$^3$ starting at $z = 90$, using $1.3 \times 10^7$ DM particles (of mass $m_{\rm DM} = 9.8 \times 10^4~\msun$) and another $1.3 \times 10^7$ SPH particles (of mass $m_{\rm gas} = 2 \times 10^4~\msun$). The gravitational softening of all particles is 0.12 physical kpc for $0 \le z \le 9$ and $1.2/(1+z)$ physical kpc for $z > 9$.

Eris2k is part of the family of Eris simulations, which were built to simulate the cosmological build-up of a local Milky Way-sized galaxy and differ only in how radiative cooling and stellar/black hole models are implemented.

In Eris2k, we assume a uniform extragalactic UV background \citep[][]{Haardt_Madau_2012} and calculate the cooling rates either in non-equilibrium (in the case of H and He) or in photoionization equilibrium (for the first 30 elements in the periodic table), using pre-computed tables made with {\textsc{cloudy}} \citep[][]{Ferland_et_al_1998}, following the model of \citet{Shen_et_al_2010,Shen_et_al_2013}.

Gas particles are allowed to form stellar particles when $\rho_{\rm gas} > 10^2 \,m_{\rm H}$~g~cm$^{-3}$, $T_{\rm gas} < 10^4$~K, and the local gas overdensity is $>$2.63. When such requirements are met, gas particles are stochastically selected such that ${\rm d}M_*/{\rm d}t = \epsilon_* M_{\rm gas} /t_{\rm dyn}$, where $M_*$ and $M_{\rm gas}$ are the mass of stars and gas involved, respectively $\epsilon_{*} = 0.1$ is the star formation efficiency, and $t_{\rm dyn}$ is the local dynamical time \citep[][]{Stinson_et_al_2006}. At each star formation event, the newly formed stellar particle (of mass $\sim6 \times 10^3~\msun$) represents a stellar population which covers the entire initial mass function by \citet{Kroupa_2001}. Metals and thermal energy are (turbulently) diffused according to the model described in \citet{Wadsley_et_al_2008} and \citet{Shen_et_al_2010}.

When a massive star (of mass between 8 and $40~\msun$) explodes as a supernova (SN), mass, iron, and oxygen are injected into the surrounding gas, together with $E_{\rm SN} = 10^{51}$~erg, which is deposited as thermal energy, according to the ``blastwave model'' of \citet{Stinson_et_al_2006}. In the case of SNae type II, radiative cooling is disabled for twice the survival time of the hot low-density shell of the SN \citep[][]{McKee_Ostriker_1977}.

The simulation results in the formation of a disc galaxy which, at $z = 0.3$, has a gaseous, a DM, and a stellar component of $1.7 \times 10^{10}$, $1.2 \times 10^{11}$, and  $3.8 \times 10^{10}~\msun$, respectively (evaluated inside a sphere of radius $20$ physical kpc and centred in the centre of mass of the galaxy). The \citet{Kron_1980} radius of the system at the same redshift is equal to 3.3~kpc.


\section{Results}\label{sec:results}

\begin{figure}
\vspace{-18pt}
\includegraphics[width=0.47\textwidth]{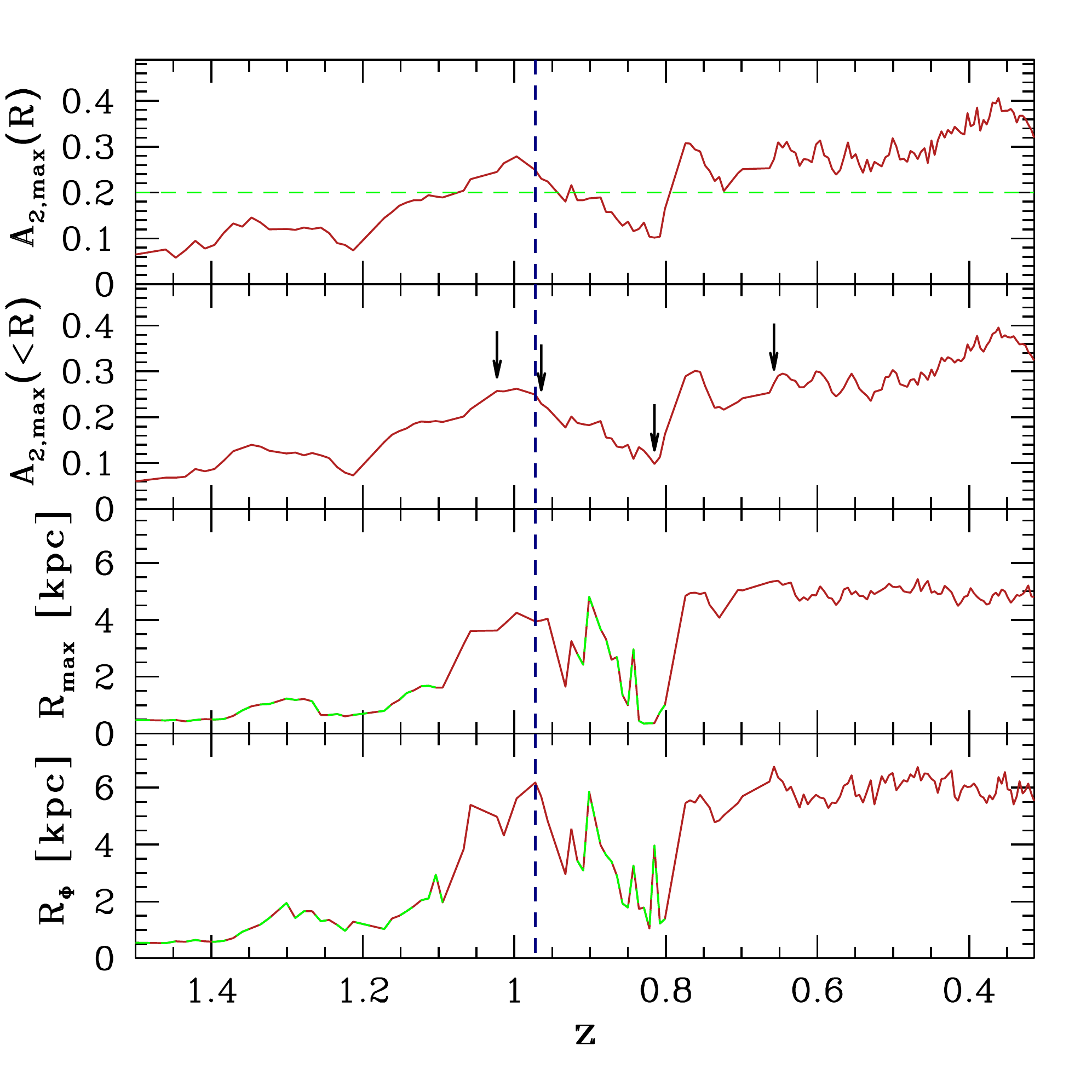}
\vspace{-10pt}
\caption{Redshift evolution of the bar properties. From top to bottom: the local intensity $A_{2, \rm{max}}(R)$ of the bar, its cumulative intensity $A_{2, \rm{max}}(<R)$, the radius $R_{\rm{max}}$ at which the cumulative intensity peaks, and the bar length $R_{\Phi}$ (see text for details). The horizontal, dashed green line in the uppermost panel marks the threshold $A_{2, \rm{max}}(R) = 0.2$. The sudden decrease in both the bar-intensity parameter and in the two bar length-scales is clearly observable immediately after the pericentre of the orbit of a satellite (a flyby), marked with a vertical, dashed blue line at $z \sim 1$. The bar length-scales estimated when $A_{2, \rm{max}}(R) < 0.2$ (i.e. when a strong bar is not present) are highlighted with green dashes. The four black arrows refer to the snapshots pictured in Figure~\ref{fig:snapshots} as notable stages in the bar evolutionary history.}
\label{fig:A2}
\end{figure}

We check for the appearance of a stellar bar and for the evolution of its strength by performing a two-dimensional Fourier analysis of the face-on stellar surface density. Following \citet{Athanassoula_Misiriotis_2002} and \citet{Valenzuela_Klypin_2003}, we define the bar strength as the ratio between the normalization of the second term in the Fourier decomposition (describing $m = 2$ modes such as bars) and the normalization of the zeroeth-order term (referring to the axisymmetric background),

\begin{equation}
A_{2}(R) = \frac{\left|\sum_{j}{m_{j}e^{2i\theta_{j}}}\right|}{\sum_{j}m_{j}} ,
\label{eq:A2}
\end{equation}

\noindent \\where $m_j$ is the mass of the $j$th particle, $\theta_{j}$ its angular coordinate in the galactic plane, and the sum is performed over all particles within a shell around the cylindrical radius $R$.\footnote{Operatively, $A_{2}(R)$ is evaluated within $R=12$ kpc by dividing the stellar disc in linearly spaced cylindrical annuli of height 2~kpc.} The strength of the bar is then taken to be $A_{\rm 2,max}(R)$, defined as the maximum of $A_{2}(R)$ at any given time.

\begin{figure*}
\includegraphics[width=0.98\textwidth]{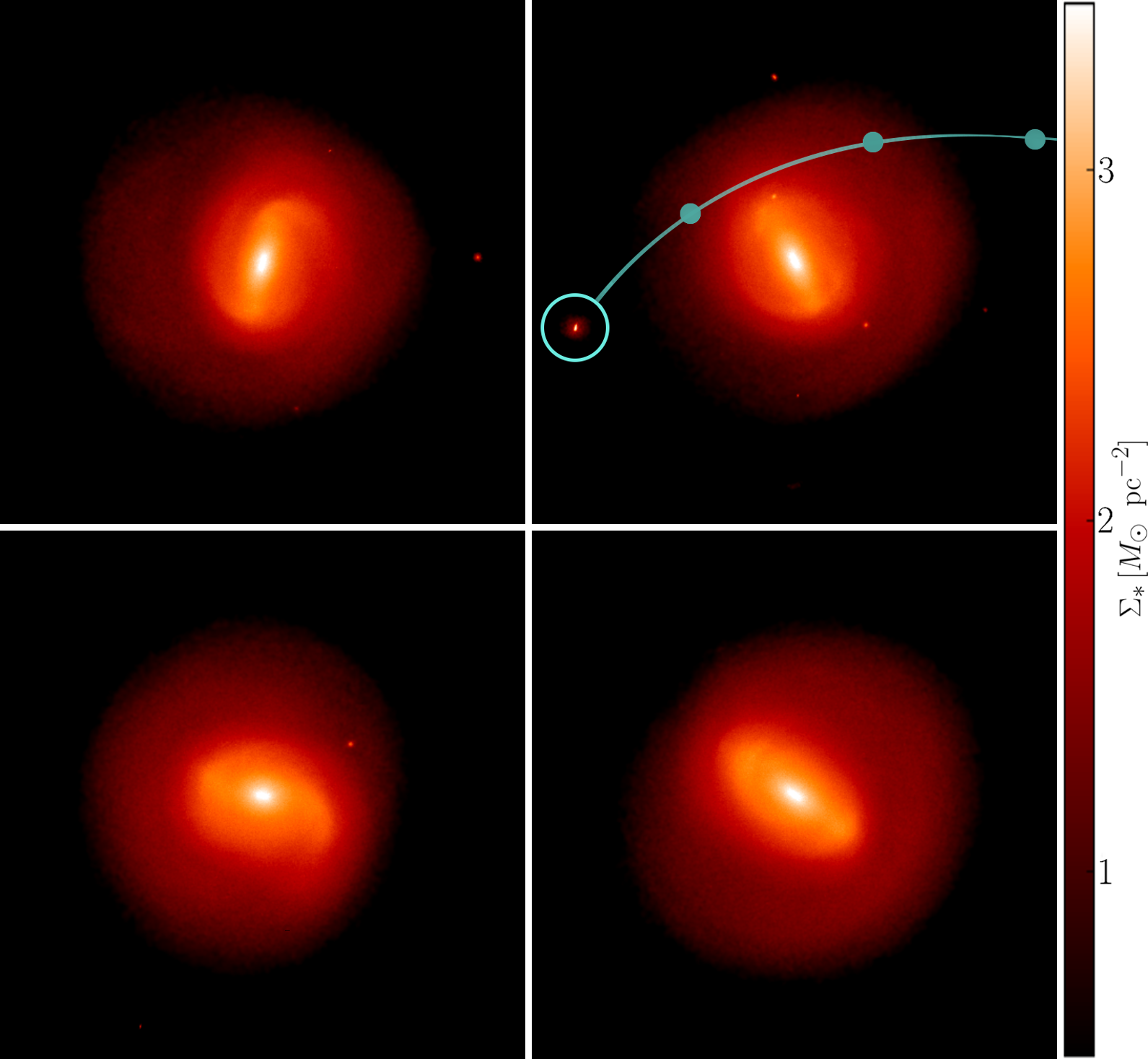}
\caption{Stellar surface density maps (viewed face-on) of the main galaxy at four different redshifts, also marked in Figure~\ref{fig:A2} (black arrows). The upper left panel refers to $z = 1.023$, before the flyby pericentre. The upper right panel ($z = 0.964$) shows the satellite (highlighted with the large circle) soon after its pericentre passage ($z = 0.972$). The positions of the satellite in the previous three snapshots (created every 33~Myr) are highlighted with smaller filled circles. The lower left panel shows the galaxy at $z = 0.815$, when a clear bar is not discernible (in response to the tidal perturbation), whereas the lower right panel shows a later snapshot ($z = 0.658$), after the bar has re-assembled.}
\label{fig:snapshots}
\end{figure*}

In addition, we compute a cumulative proxy for the bar intensity, $A_{2}(<R)$ [and its maximum $A_{\rm 2,max}(<R)$], differing from $A_{2}(R)$ only by considering all the particles within a given radius $R$. This second proxy is less precise in determining the maximum strength of the bar but, since it keeps growing with $R$ as long as a bar-like structure is present, the radius $R_{\rm max}$ at which $A_{2}(<R)$ peaks can be used as an estimate for the extent of the bar \citep[see][]{Spinoso_et_al_2017, Zana_et_al_2018}.

An alternative estimate for the bar length is the radius $R_{\Phi}$ at which the phase of the $m = 2$ mode,

\begin{equation}
\Phi(R) \equiv \frac{1}{2} \arctan \left[ \frac{\sum_{j}{m_{j}\sin(2\theta_{j})}}{\sum_{j}{m_{j}\cos(2\theta_{j})}} \right] ,
\label{eq:phase}
\end{equation}

\noindent \\deviates from the approximately constant value defining the bar orientation (here we follow a procedure described in \citealt{Athanassoula_Misiriotis_2002} with the modifications discussed in \citealt{Zana_et_al_2018}).\footnote{In detail, a tolerance $\Delta \Phi = \arcsin(0.15)$ is used and the phase of the bar is evaluated at $R_{\rm max}$.}

The evolution of the two strength parameters and of the two bar lengths for $z\lsim 1.5$ is shown in Figure~\ref{fig:A2}. A strong bar [i.e. a bar having $A_{\rm 2,max}(R)>0.2$] forms at $z \simeq 1.1$, rapidly increasing its size up to a considerable fraction of the stellar disc: at $z \sim 1$, $R_{\rm max} \gsim 4$~kpc and $R_{\Phi} \simeq 6$~kpc, whereas the Kron radius is $\sim$3~kpc. The bar keeps growing in intensity with an approximately constant length until the end of the run at $z = 0.3$, with the only exception of a transient weakening at about $0.95\gsim z \gsim 0.8$. The weakening corresponds to a lower coherence of the bar structure, making the measurement of the bar extent in this phase more difficult and less significant (this is marked in Figure~\ref{fig:A2} by the dashed green-red lines). However, a clear decrease in the bar extent is still observable in the same redshift interval.
 
The physical trigger of the transient fading of the bar is the gravitational perturbation of a satellite galaxy undergoing its pericentre at $z \sim 1$ on a prograde orbit, well within the stellar disc ($R_{\rm peri} \simeq 6.5$~kpc). The apocentre-to-pericentre ratio of 6:1 agrees very well with the cosmologically expected value derived by \citet{Ghigna_et_al_1998}. The total (baryonic plus DM) mass of the perturber immediately before the pericentric passage (evaluated at a separation $\sim$20~kpc) is $M_{\rm sat} \simeq 1.1 \times 10^8 \msun$, whereas the main system has a total mass $M_{\rm gal} \simeq 1.5 \times 10^{11} \msun$. It is important to note that the stellar mass ratio, $q_{*} \simeq 4.5 \times 10^{-3}$, albeit low, is larger than the total mass ratio since the perturber is not DM-dominated, contrary to the main galaxy.\footnote{Note that, while the satellite's DM halo is strongly affected by tidal stripping, almost its whole stellar component is preserved during the flyby.} After the bar has formed, the aforementioned perturber is the most massive amongst those objects with similar $R_{\rm peri}$.\footnote{A perturber of similar mass and similar $R_{\rm peri}$ is observed at $z \sim 0.35$, concurrent with a small bar-strength decrease. However, in this case, we are not able to confirm if the bar disappears (either permanently or temporarily), since the run ends at $z = 0.3$.}

The effect of the satellite's passage on to the main galaxy is shown in Figure~\ref{fig:snapshots} for the four redshifts marked with arrows in Figure~\ref{fig:A2}. The upper left panel shows the galactic morphology after the bar formation, but before the satellite's pericentre. A stage close to the pericentric passage is shown in the upper right panel, with the position of the satellite highlighted with a large circle and its position in the three previous snapshots marked with smaller filled circles. The perturbation alters the orbits of the bar's stars, and the bar loses its coherence (see the lower left panel) for $\sim$900~Myr.

The bar reforms with a size and strength very similar to those before the encounter, and this is observable in the lower-right panel. This is because the potential of the galaxy is almost unperturbed by the satellite passage, as shown in Figure~\ref{fig:prec} through the profiles of the orbital $\Omega$ and precession $\Omega-\kappa/2$ frequencies\footnote{$\kappa=\sqrt{\left(\frac{\rm{d}^2\phi}{{\rm d}R^2}+\frac{3}{R}\frac{{\rm d}\phi}{{\rm d}R}\right)}_{R}$ is the epicyclic frequency, i.e. the frequency of small radial oscillations in the potential $\phi$. As a consequence, $\Omega-\kappa/2$ is the precession frequency of an otherwise close orbit with two radial oscillations per revolution.} before the strong interaction ($z_1 = 1.023$; upper panel in Figure~\ref{fig:prec} and upper-left panel in Figure~\ref{fig:snapshots}) and during the faded-bar phase ($z_2 = 0.815$; middle panel in Figure~\ref{fig:prec} and lower-left panel in Figure~\ref{fig:snapshots}). We stress that $\Omega-\kappa/2$ is particularly suited for this test, since its shape is extremely sensitive to small variations in the potential and, for this reason, it is used to determine whether a non-axisymmetric perturbation has Lindblad resonances.\footnote{For instance, the $\Omega-\kappa/2$ profile has been also used to constrain the mass of central massive black holes even when their influence radii are not resolved; see, e.g. \citet{Combes_et_al_2014}.} The potential profile remains unchanged because the perturbation does not lead to a  significant variation in the mass distribution of the disc. We also verified that a central mass concentration does not grow during the flyby passage.\footnote{The growth of a central mass concentration would strongly affect the precession frequencies in the inner disc region, leading to the bar destruction \citep[see, e.g.][]{Shen_Sellwood_2004, Kormendy_2013}.}

\subsection{Comparison with previous works}

This work strengthen the results of previous analysis performed on isolated simulations.
\cite{Gerin_et_al_1990} described the change in bar strength and pattern speed of a strongly barred galaxy after a close encounter in isolated three-dimensional models.
They found that the interaction causes a transient increase (or decrease) of the bar strength and a corresponding decrease (or increase) of the pattern speed, depending on the angle $\alpha \equiv \phi_{{\rm bar}}-\phi_{{\rm sat}}$, where $\phi_{{\rm bar}}$ is the phase of the bar and $\phi_{{\rm sat}}$ is the angle of the perturber at the moment of its pericentre, in the reference frame of the centre of mass of the main galaxy. \footnote{When $\alpha$ is positive (negative), the perturber is behind (ahead of) the bar; when $\alpha = \pm90^\circ$, the effect of the satellite is null, due to simmetry reasons.}
According to \cite{Gerin_et_al_1990}, when $\alpha$ is positive, the companion extracts angular momentum from the particles at the terminal edges of the bar. For this reason, they move on more eccentric orbits and contribute to increase the length and the strength of the bar. As a consequence, the bar becomes slower, since the corotation migrates toward the outer part of the disc.\footnote{The opposite effect arises for negative angles.}
In Eris2k, it is hard to evaluate the precise moment of the pericentre, given the temporal resolution of 33 Myr (the cosmological nature of the simulation prevented a more frequent sampling of the galactic history).
In the snapshot with the shortest distance between the perturber and the centre of mass of the main system (the one we adopt to compute $R_{\rm peri}$), $\alpha = 82^\circ$, which is very close to the neutral angle of $\pm90^\circ$, leaving a considerable uncertainity on the interpretation of the interaction.
On the other hand, in the snapshots immediately before and after, the angles are equal to $\alpha = -54^\circ$ and $\alpha = -79^\circ$, respectively, and these results are compatible with the conclusion of \cite{Gerin_et_al_1990}.
Moreover, our results are not in disagreement with a slight increase of the bar angular velocity, but the trend is not obvious at all, as the variation is small and we do not have a reference simulation of the galaxy evolving in isolation in order to evaluate the phase difference \citep[as performed in][]{Gerin_et_al_1990}.
However, the mass ratios $M_{{\rm gal}}/M_{{\rm sat}}$ they used in their study (0.5 and 1) are far greater than ours ($7\times10^{-4}$), whereas our pericentre is similar to the values they investigated. 
It follows that, in Eris2k, the perturbation is minimal with respect to those studied in \cite{Gerin_et_al_1990}, but we found that this is sufficient to lower the bar strength parameter (see Figure~\ref{fig:A2}) by about 70 per cent of its own value before the satellite pericentre.

It is interesting to discuss how the statistical study of \cite{Peschken_Lokas_2018} is related to our work.
The authors analysed, in the Illustris-1 simulation, a sample of 121 massive galaxies which underwent a close encounter during their evolutionary history and found a trace of that interaction in the evolution of the bar strength parameter.
They argued that the effect of the perturbation on the structure would depend both on the orbital angle (i.e. whether the approaching satellite is on a prograde or retrograde orbit) and on the strength of the perturbation.
On average, the overall effect of a weak tidal perturbation in their sample is to reduce the strength of the bar. 
In order to provide a quantitative comparison of our results with respect to this work, we computed the angle $\theta$ between the plane of the orbit of the satellite galaxy and the disc of the main system, along with the Elmegreen tidal strength parameter $S$ \citep{Elmegreen_et_al_1991}.
Our perturbation is very weak, given the really small mass ratio, yielding $S = 4.8\times10^{-3}$. The orbit is almost completely coplanar with the primary galaxy disc, in close proximity of the pericentre, with $\cos(\theta) = 0.96$. These values are in agreement with what the authors found in the Illustris run, confirming the scenario of small perturbations that work against the bar growth, independent of their orbital angular momentum.

\section{Conclusions}\label{sec:conclusions}

\begin{figure}
\vspace{-10pt}
\includegraphics[width=0.47\textwidth]{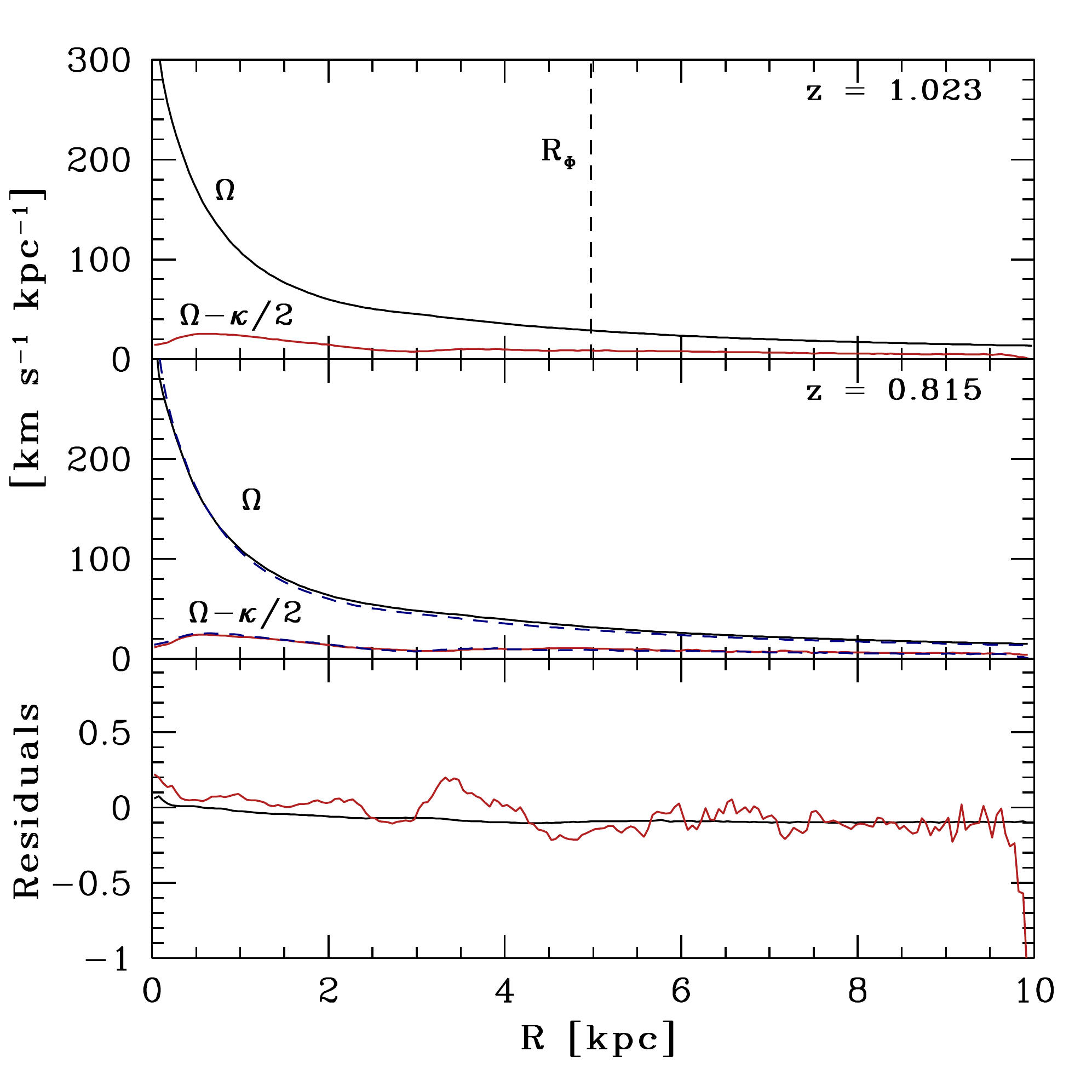}
\caption{Frequency plot for the main galaxy. Upper panel: angular frequency $\Omega$ (solid black curve) and precession frequency $\Omega-\kappa/2$ (solid red curve) at $z_1 = 1.023$, before the satellite's pericentric passage (see the upper left panel of Fig.~\ref{fig:snapshots}). The vertical, dashed black line marks the length $R_{\Phi}$ of the bar at that time. Middle panel: angular frequency $\Omega$ (solid black curve) and precession frequency $\Omega-\kappa/2$ (solid red curve) at $z_2 = 0.815$, after the satellite's pericentric passage (see the lower left panel of Fig.~\ref{fig:snapshots}). For an easier comparison, we also show (dashed blue curves) the $\Omega$ and $\Omega-\kappa/2$ values evaluated at $z_1$. Lower panel: relative variations of $\Omega$ (black curve), defined as $[\Omega(z_1)-\Omega(z_2)]/\Omega(z_1)$, and of $\Omega-\kappa/2$ (using the same definition; red line).}
\vspace{-10pt}
\label{fig:prec}
\end{figure}

We investigated the effect of the interactions between a growing spiral galaxy and its satellites in the cosmological zoom-in simulation Eris2k, focussing on the impact that these interactions have on the persistence of the bar in the main galaxy. We identified the main interaction after the bar appearance as an unequal-mass satellite ($q_{*} \simeq 4.5 \times 10^{-3}$) which undergoes a close pericentric passage ($R_{\rm peri} \simeq 6.5$~kpc, of the order of the bar length). 

The perturbation shuffles the orbits of the stars which make up the bar, weakening and shortening the bar for about $900$~Myr. The bar then reforms with strength and length comparable to those before the interaction. The reason for such bar resilience is due to the fact that the profile of the gravitational potential of the main galaxy and, as a consequence, that of its orbital and precession frequencies are not significantly affected by the perturbation. 

This study thus provides a further indication that close unequal-mass encounters (most common during the cosmological evolution of structures) have a small impact on the internal evolution of field disc galaxies, where bars form spontaneously from small seed deviations from axisymmetry \citep[e.g.][]{Moetazedian_et_al_2017,Zana_et_al_2018}.

However, we stress that, due to the non-negligible duration of the bar-less period, a fraction of observed spiral galaxies classified as non-barred or weakly barred could be prone to bar formation. These galaxies could have hosted a strong bar in the (recent) past and currently be undergoing a bar-regrowth phase after a close flyby. We further speculate that, during such phase, a weak deviation from axisymmetry, in the form of lenses/ovals could be observed\footnote{A central oval/lens is observable in the lower-left panel of Figure~\ref{fig:snapshots}, but we caution the reader that a more detailed analysis (including the modelling of mock images) is needed to prove the actual observability of this morphological structure.} and misinterpreted as the sign of a ``bar suicide'' (see \citealt{Kormendy_2013} and references therein), if no analysis of the host potential is performed. The frequency of close flyby occurrences and their relative impact in the fraction of lensed/ansaed spirals cannot be assessed with a single cosmological zoom-in simulation. Moreover, we noticed that the bulge-to-disc ratio, as measured through the stellar surface density fitting, decreases when the bar is suppressed, similarly to what observed during major mergers in \citet{Guedes_et_al_2013}. A more detailed analysis is required and will be the focus of a follow-up study.

We saw that, due to the very unequal-mass ratio of the encounter, the disc potential remained basically unperturbed by the satellite passage (this is clearly shown in Figure~\ref{fig:prec}). As a consequence, the current case does not result in the definitive disintegration of the non-axisymmetric structure (also known as the bar suicide). The cause for the ``apparent death'' of the bar in Eris2k is a minor and temporary energy exchange induced by the close tidal interaction.
Even a small amount of energy is sufficient to azimuthally perturb the orbits in an almost-axisymmetric disc potential and, for this reason, the general effect of the flyby in the upper right panel of Figure~\ref{fig:snapshots} is to undermine the coherence of the orbits which contribute to the body of the bar, blurring its structural integrity.
Once the perturbation has ceased, the self-gravity of the bar relic resumes the bar overdensity by slowly dissipating the energy of the encounter.
Therefore, the bar is restored, but initially it appears to be puffier and less defined, as shown in the lower left panel of Figure~\ref{fig:snapshots}.

An analogous fate has been observed in previous work. Depending on the development stage of the bar, either its growth can appear delayed, as in \cite{Zana_et_al_2018}, or its whole structure can be destroyed (if the bar is fully grown and established) for a limited period, as shown in this work. The recurrent weakening of the bar due to encounters with satellites could be more important, and eventually critical, at higher redshift, as the interaction rate is supposed to increase.

\vspace{-0.4cm}
\section*{Acknowledgements}
The Authors would like to thank the Referee for the helpful suggestions that improved the quality of the manuscript. PRC acknowledges support by the Tomalla foundation.

\scalefont{0.94}
\setlength{\bibhang}{1.6em}
\setlength\labelwidth{0.0em}
\bibliographystyle{mnras}
\bibliography{bar_resilience}
\normalsize

\bsp	
\label{lastpage}
\end{document}